\documentclass[twocolumn,showpacs,preprintnumbers,amsmath,amssymb]{revtex4}

\usepackage{graphicx}
\usepackage{dcolumn}
\usepackage{bm}

\newcommand{\bq}{\begin{equation}}
\newcommand{\eq}{\end{equation}}
\newcommand{\bqa}{\begin{eqnarray}}
\newcommand{\eqa}{\end{eqnarray}}
\newcommand{\nn}{\nonumber \\}

\def\be     {\begin{equation}}
\def\ee     {\end{equation}}
\def\bea        {\begin{eqnarray}}
\def\eea        {\end{eqnarray}}
\def\bnn    {\begin{eqnarray*}}
\def\enn    {\end{eqnarray*}}

\begin{document}

\title{Evidence of electron fractionalization in the Hall
coefficient at Mott criticality}
\author{Ki-Seok Kim}
\affiliation{Asia Pacific Center for Theoretical Physics, Pohang, Gyeongbuk 790-784, Republic of Korea \\
Department of Physics, POSTECH, Pohang, Gyeongbuk 790-784, Korea}

\date{\today}

\begin{abstract}
%
%
Hall coefficient implies the mechanism for reconstruction of a
Fermi surface, distinguishing competing scenarios for Mott
criticality such as electron fractionalization, dynamical
mean-field theory, and metal-insulator transition driven by
symmetry breaking. We find that electron fractionalization leaves
a signature for the Hall coefficient at Mott criticality in two
dimensions, a unique feature differentiated from other theories.
We evaluate the Hall coefficient based on the quantum Boltzman
equation approach, guaranteeing gauge invariance in both
longitudinal and transverse transport coefficients.
\end{abstract}


\maketitle

\section{Introduction}

Since reconstruction of Fermi surfaces has been discussed in the
context of high T$_{c}$ cuprates
\cite{FS_cuprates_View,FS_cuprates_Exp_Original,FS_cuprates_Exps},
it is considered to play an essential role in understanding the
mechanism of metal-insulator transitions \cite{Review_MIT}. Its
information is reflected in the Hall coefficient
\cite{Hall_Mott,Hall_HF1,Hall_HF2}, expected to distinguish
competing scenarios for Mott criticality based on electron
fractionalization \cite{GT_MIT}, dynamical mean-field theory
(DMFT) \cite{DMFT_MIT}, and metal-insulator transition driven by
symmetry breaking (SB-MIT), respectively. Indeed, the
Fermi-surface reconfiguration has been regarded as one of the
central problems in heavy-fermion quantum criticality
\cite{HF_Review}, claimed to differentiate competing theories
\cite{Kim_GR,Kim_WF,Kim_Seebeck,Kim_Two_Energy} such as the Kondo
breakdown scenario \cite{Paul_KB,Pepin_KB}, the local quantum
criticality ansatz \cite{Si_LQCP}, and the spin-density-wave
theory \cite{HMM}.

In this paper we propose evidence of electron fractionalization at
Mott criticality from the Hall coefficient. An essential feature
of the fractionalizaion scenario is appearance of novel charge
fluctuations (holons) near the Mott critical point, giving rise to
an interesting metallic state in two dimensions. Such charge
fluctuations turn out to contribute to the Hall coefficient
additionally on top of the Fermi-liquid value given by fermion
excitations (spinons). As a result, we uncover two-step jumps in
the Hall coefficient at the two dimensional Mott critical point.
On the other hand, both DMFT and SB-MIT allow only one-step jump
from the Fermi-liquid value to infinity at the Mott critical
point. See Fig. 1.
%
%

\begin{figure}
\includegraphics[width=0.48\textwidth]{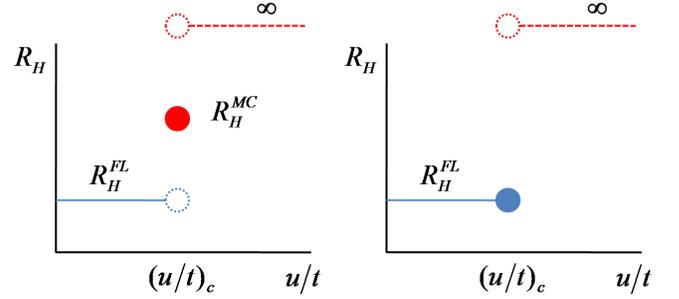}
\caption{A schematic diagram for the Hall coefficient. Electron
fractionalization gives rise to two-step jumps in the Hall
coefficient at the Mott critical point in two spatial dimensions
(left), where novel charge fluctuations (holons) appear to
contribute to the Hall coefficient in addition to fermions
(spinons). On the other hand, either DMFT or SB-MIT allows only
one-step jump (right). $FL$ means Fermi liquid and $MC$ denotes
Mott critical point. } \label{fig1}
\end{figure}

A scaling theory for fermions has been proposed near Mott
criticality based on an ansatz of electron fractionalization,
where the scaling expression of the spectral function is
determined by both the dynamical critical exponent and the
anomalous dimension of fermions \cite{Senthil_Scaling}. Although
this scaling theory serves a convenient framework for quantum
criticality in thermodynamics and transport, it is difficult to
find such critical exponents in the Fermi-surface problem
\cite{SSL,Max}, where they are introduced phenomenologically. This
difficulty does not arise in the Hall coefficient, not
renormalized by strong correlations, thus suggesting an
undisputable evidence for electron fractionalization as the
mechanism of the Fermi-surface reconstruction.

\section{A signature of electron fractionalization in the Hall
coefficient}

We start from the Hubbard model \bqa && H = - t \sum_{\langle i j
\rangle} (c_{i\sigma}^{\dagger} c_{j\sigma} + H.c.) + u \sum_{i}
n_{i\uparrow} n_{i\downarrow} . \eqa $c_{i\sigma}$ is an electron
field, and $n_{i\sigma} = c_{i\sigma}^{\dagger}c_{i\sigma}$ is its
density. $t$ is a hopping parameter, and $u$ is an interaction
strength at site $i$. In this study we focus on the case of two
spatial dimensions.

A way to describe electron fractionalization is to extract out
collective charge dynamics explicitly from correlated electrons,
referred as the slave-rotor representation
\cite{Florens,Kim_Graphene}. Such charge fluctuations are
identified with zero sound modes in the case of short range
interactions while plasmon modes in the case of long range
interactions. Actually, one can check that the dispersion of the
rotor variable is the same as that of such collective charge
excitations. The Mott transition is described by gapping of rotor
excitations in the slave-rotor theory.

The slave-rotor theory is based on decomposition of an electron
field as follows $c_{i\sigma} = e^{-i\theta_{i}} f_{i\sigma}$,
where $e^{-i\theta_{i}}$ is a holon field to carry an electron
charge, and $f_{i\sigma}$ is a spinon field to do an electron
spin. Resorting to this fractionalization scenario, one can
rewrite the Hubbard model in terms of holons and spinons,
interacting via gauge fluctuations. Gauge-field excitations are
associated with collective spin fluctuations, differentiated from
those in the spin-fermion model \cite{HMM}. Performing the
continuum approximation, we reach the following expression for the
Mott transition \bqa && {\cal L} = f_{\boldsymbol{x}
\sigma}^{\dagger} (\partial_{\tau} - \mu - i
\varphi_{\boldsymbol{x} }) f_{\boldsymbol{x} \sigma} +
\frac{1}{2m_{f}} |(\boldsymbol{\nabla} - i
\boldsymbol{a}_{\boldsymbol{x}}) f_{\boldsymbol{x} \sigma}|^{2}
\nn && + \frac{1}{2 u} (\partial_{\tau}\theta_{\boldsymbol{x} } -
\varphi_{\boldsymbol{x} })^{2} + \frac{1}{2m_{b}}
(\boldsymbol{\nabla}\theta_{\boldsymbol{x} } -
\boldsymbol{a}_{\boldsymbol{x}})^{2} + \frac{1}{4g^{2}} f_{\mu\nu}
f_{\mu\nu} , \nn \eqa where the last term with the field strength
tensor $f_{\mu\nu} = \partial_{\mu} a_{\nu} - \partial_{\nu}
a_{\mu}$ describes the Maxwell dynamics of gauge fluctuations, and
the gauge-matter coupling constant is denoted as $g$. $m_{f(b)}$
is the band mass of spinons (holons). A detailed derivation of
this procedure can be found in Refs. \cite{Florens,Kim_Graphene}.

This effective U(1) gauge theory has been discussed in the
Eliashberg framework, where self-energy corrections of spinons,
holons, and gauge fields are incorporated self-consistently, but
vertex corrections are neglected \cite{Senthil_MIT1,Senthil_MIT2}.
Integrating over both spinons and holons, one finds an effective
action for gauge fluctuations \bqa && \mathcal{S}_{G} =
\frac{1}{\beta} \sum_{i\Omega} \int \frac{d^{2}
\boldsymbol{q}}{(2\pi)^{2}} \frac{1}{2} \Bigl( \gamma
\frac{|\Omega|}{q} + \chi q^{2} + \sigma_{\theta} \sqrt{\Omega^{2}
+ q^{2}} \Bigr) \nn && \boldsymbol{a}_{i}(\boldsymbol{q},i\Omega)
\Bigl( \delta_{ij} - \frac{q_{i}q_{j}}{q^{2}}\Bigr)
\boldsymbol{a}_{j}(-\boldsymbol{q},-i\Omega) , \eqa where
multi-scale quantum criticality is uncovered from two dynamical
critical exponents, $z = 1$ from critical dynamics of holons and
$z = 3$ from particle-hole excitations near the Fermi surface of
spinons. $\gamma$ is the damping coefficient, proportional to the
density of states of spinons, and $\chi$ is the diamagnetic
susceptibility of spinons. $\sigma_{\theta}$ is the holon
conductivity. Resorting to this gauge propagator, one can evaluate
self-energy corrections of spinons and holons self-consistently in
each regime. It was shown that the holon self-energy due to gauge
fluctuations is not relevant, implying that the Mott transition
falls into the 3D-XY universality class
\cite{Senthil_MIT1,Senthil_MIT2}. On the other hand, the spinon
self-energy gives rise to scaling for thermodynamics and transport
in each regime.

In order to evaluate both longitudinal and transverse transport
coefficients, we resort to the quantum Boltzman equation approach
\cite{Mahan}, given by
\begin{widetext}
\bqa && \frac{e}{c} \{\boldsymbol{v}_{k}^{f} +
\boldsymbol{\partial}_{k} \Re \Sigma_{ret}^{f}(k,\omega) \} \cdot
(\boldsymbol{\mathcal{B}} \times \boldsymbol{\partial}_{k})
G_{f}^{<}(k,\omega) \nn && + e \boldsymbol{\mathcal{E}} \cdot
[\{\boldsymbol{v}_{k}^{f} + \boldsymbol{\partial}_{k} \Re
\Sigma_{ret}^{f}(k,\omega)\} \Gamma_{f}(k,\omega) + \{\omega -
\xi_{k}^{f} - \Re \Sigma_{ret}^{f}(k,\omega)\}
\boldsymbol{\partial}_{k} \Gamma_{f}(k,\omega)] [
\partial_{\omega} n_{f}(\omega)]
[A_{f}(k,\omega)]^{2} = I_{coll}^{f}(k,\omega) , \nn &&
I_{coll}^{f}(k,\omega) = i [2 \Gamma_{f}(k,\omega)
G^{<}_{f}(k,\omega) - \Sigma^{<}_{f}(k,\omega) A_{f}(k,\omega)]
\eqa for spinons and \bqa && \frac{e}{c} \{\boldsymbol{v}_{k}^{b}
+ \boldsymbol{\partial}_{k} \Re \Sigma_{ret}^{b}(k,\omega) \}
\cdot [(\boldsymbol{B} + \boldsymbol{\mathcal{B}}) \times
\boldsymbol{\partial}_{k}] G_{b}^{<}(k,\omega) \nn && + e
(\boldsymbol{E} + \boldsymbol{\mathcal{E}}) \cdot
[\{\boldsymbol{v}_{k}^{b} + \boldsymbol{\partial}_{k} \Re
\Sigma_{ret}^{b}(k,\omega)\} \Gamma_{b}(k,\omega) + \{\omega -
\xi_{k}^{b} - \Re \Sigma_{ret}^{b}(k,\omega)\}
\boldsymbol{\partial}_{k} \Gamma_{b}(k,\omega)] [\partial_{\omega}
n_{b}(\omega)] [A_{b}(k,\omega)]^{2} = I_{coll}^{b}(k,\omega) ,
\nn && I_{coll}^{b}(k,\omega) = i [2 \Gamma_{b}(k,\omega)
G^{<}_{b}(k,\omega) - \Sigma^{<}_{b}(k,\omega) A_{b}(k,\omega)]
\eqa
\end{widetext} for holons, where $G_{f(b)}^{<}(k,\omega)$
and $\Sigma_{f(b)}^{<}(k,\omega)$ are lesser Green's function and
self-energy of spinons (holons), respectively, and
$A_{f(b)}(k,\omega)$, $\Gamma_{f(b)}(k,\omega)$, and $\Re
\Sigma_{ret}^{f(b)}(k,\omega)$ are imaginary parts of retarded
Green's function, self-energy, and the real part of the retarded
self-energy, respectively. $\boldsymbol{v}_{k}^{f(b)}$ and
$\xi_{k}^{f(b)}$ are velocity and dispersion of spinons (holons),
respectively. $n_{f(b)}(\omega)$ is the Fermi-Dirac
(Bose-Einstein) distribution function. $\boldsymbol{E}$ and
$\boldsymbol{B}$ are applied electric and magnetic fields while
$\boldsymbol{\mathcal{E}}$ and $\boldsymbol{\mathcal{B}}$ are
internal fields related with fractionalization. Since spinons do
not carry an electric charge in our assignment, they couple to
internal fields only in a gauge invariant way. Equations (4) and
(5) are derived in appendix A.

Inelastic scattering with critical fluctuations gives rise to the
collision term of the right-hand-side, where each lesser
self-energy is given by \begin{widetext} \bqa &&
\Sigma_{f}^{<}(k,\omega) = \sum_{q} \int_{0}^{\infty}
\frac{d\nu}{\pi} \Bigl| \frac{k\times\hat{q}}{m_{f}} \Bigr|^{2}
\Im D_{a}(q,\nu) [\{n_{b}(\nu) + 1\} G_{f}^{<}(k+q,\omega+\nu) +
n_{b}(\nu) G_{f}^{<}(k+q,\omega-\nu)] , \nn &&
\Sigma_{b}^{<}(k,\omega) = \sum_{q} \int_{0}^{\infty}
\frac{d\nu}{\pi} \Bigl( \Bigl| \frac{k\times\hat{q}}{m_{b}}
\Bigr|^{2} \Im D_{a}(q,\nu) + \Im D_{\lambda}(q,\nu) \Bigr)
[\{n_{b}(\nu) + 1\} G_{b}^{<}(k+q,\omega+\nu) + n_{b}(\nu)
G_{b}^{<}(k+q,\omega-\nu)] . \nn \eqa
\end{widetext} $\Im D_{a}(q,\nu)$ is the spectral function of
the gauge propagator, given by $z = 3$ in the paramagnetic
insulating state and $z = 2$ at the Mott critical point. $\Im
D_{\lambda}(q,\nu)$ is the spectral function of the $\lambda$
propagator, where excitations of $\lambda$ correspond to potential
fluctuations for holons in the nonlinear $\sigma$ model
description \cite{Lambda_Dynamics}. Such fluctuations are
essential, the Mott transition belonging to the XY universality
class \cite{Senthil_MIT1,Senthil_MIT2}.
%
%

These Boltzman equations are simplified further in the Eliashberg
approximation, where self-energy corrections do not depend on
momentum. Inserting the lesser Green's function \bqa &&
G_{f(b)}^{<}(k,\omega) = i n_{f(b)}(\omega) A_{f(b)}(k,\omega) \nn
&& - i \Bigl( \frac{\partial n_{f(b)}(\omega)}{\partial \omega}
\Bigr) A_{f(b)}(k,\omega) \boldsymbol{v}^{f(b)}_{k} \cdot
\boldsymbol{\Lambda}_{f(b)}(k,\omega) \eqa into the quantum
Boltzman equations (4) and (5) with Eq. (6), we find so called
vertex distribution functions for spinons and holons \bqa &&
\Lambda_{f}(k_{F},\omega) = \frac{e}{2}
\frac{\tau_{tr}^{f}(k_{F},\omega)}{\tau_{sc}^{f}(k_{F},\omega)}
\frac{A_{f}(k_{F},\omega)}{1 + i \Omega_{f}
\tau_{tr}^{f}(k_{F},\omega)} \mathcal{E} , \nn &&
\Lambda_{b}(k,\omega) = \frac{e}{2}
\frac{\tau_{tr}^{b}(k,\omega)}{\tau_{sc}^{b}(k,\omega)}
\frac{A_{b}(k,\omega)}{1 + i (\Omega_{b} + \omega_{b})
\tau_{tr}^{b}(k,\omega)} ( \mathcal{E} + E ) , \nn \eqa
respectively, where $\Omega_{f(b)} = \frac{e \mathcal{B}}{m_{f(b)}
c}$ and $\omega_{f(b)} = \frac{e B}{m_{f(b)} c}$ are internal and
external cyclotron frequencies, and $\mathcal{E} = \mathcal{E}_{x}
+ i \mathcal{E}_{y}$ and $E = E_{x} + i E_{y}$ are complex fields.
$[\tau_{sc}^{f(b)}(k,\omega)]^{-1} = 2 \Gamma_{f(b)}(k,\omega)$ is
the relaxation rate and $[\tau_{tr}^{f(b)}(k,\omega)]^{-1}$ is the
scattering rate associated with the transport time, where the
former quantity is obtained from the Eliashberg calculation, and
the latter is found with vertex corrections, thus a gauge
invariant quantity. A detailed procedure on how to solve quantum
Boltzman equations is presented in appendix B.

Inserting these vertex-distribution functions into expressions for
currents, we obtain \bqa && J_{f} = \mathcal{C}_{f} e^{2}
\rho_{F}^{f} v_{F}^{f 2} \frac{\tau_{tr}^{f}(T)}{1 + i \Omega_{f}
\tau_{tr}^{f}(T)} \mathcal{E} , \nn && J_{b} = \mathcal{C}_{b}
e^{2} \frac{\tau_{tr}^{b}(T)}{1 + i (\Omega_{b} + \omega_{b})
\tau_{tr}^{b}(T)} ( \mathcal{E} + E ) , \eqa where $\rho_{F}^{f}$
and $v_{F}^{f}$ are the density of states and Fermi velocity of
spinons with a positive numerical constant $\mathcal{C}_{f}$, and
$\mathcal{C}_{b}$ is a positive number given by a function of
temperature and the ratio $u/t$. $\mathcal{C}_{b}$ is a universal
number at the Mott critical point, and it vanishes in the
paramagnetic insulator.

Internal magnetic fields are shifted as \bqa && \Omega_{f} =
\frac{\chi_{b}}{\chi_{f} + \chi_{b}} \omega_{f} , ~~~~~ \Omega_{b}
+ \omega_{b} = \frac{\chi_{f}}{\chi_{f} + \chi_{b}} \omega_{b} ,
\eqa determined from the gauge invariance \cite{Lee_Nagaosa}.
Another condition for the gauge invariance is the back-flow
constraint, \bqa && J_{f} + J_{b} = 0 , \eqa implying that the
total internal current should vanish \cite{Ioffe_Larkin}.
Then, we reach the final expression for the Hall resistivity \bqa
&& \rho_{xy}(T) = \frac{\sigma_{xy}(T)}{[\sigma_{xx}(T)]^{2} +
[\sigma_{xy}(T)]^{2}} \nn && \approx \frac{1}{e^{2}
\mathcal{C}_{b}} \frac{\chi_{f}}{\chi_{f} + \chi_{b}} \omega_{b} +
\frac{1}{e^{2} \mathcal{C}_{f} \rho_{F}^{f} v_{F}^{f 2}}
\frac{\chi_{b}}{\chi_{f} + \chi_{b}} \omega_{f} . \eqa It is
important to notice that the transport time does not appear,
implying that renormalization of the Hall conductivity due to
quantum corrections is cancelled by that of the square of the
longitudinal conductivity. A derivation is shown in appendix C.
%
%

In the Fermi liquid phase ($u < u_{c}$) holon condensation gives
rise to $\chi_{b} \rightarrow \infty$ \cite{Lee_Nagaosa}. Then, we
find $\rho_{xy}(T) \approx \frac{ \omega_{f} }{e^{2}
\mathcal{C}_{f} \rho_{F}^{f} v_{F}^{f 2}}$, which coincides with
that of the Fermi liquid state. On the other hand, $\chi_{f} \gg
\chi_{b}$ results both at the Mott critical point and in the
paramagnetic insulator. Then, we obtain $\rho_{xy}(T) \approx
\frac{ \omega_{b}}{e^{2} \mathcal{C}_{b}}$, resulting from only
holons. In the insulating phase we see $\mathcal{C}_{b}
\rightarrow 0$. See appendix D. As a result, we obtain
$\rho_{xy}(T \rightarrow 0) \rightarrow \infty$. However, it is
not the case at the critical point, where the critical dynamics of
holons is characterized by their own conductivity.
$\mathcal{C}_{b}$ is given by a constant, leading the Hall
coefficient at the Mott critical point to differ from the
Fermi-liquid value.

Integrating over spinon excitations, we obtain an effective field
theory for critical holon dynamics with the $z = 2$ gauge dynamics
given by Eq. (3) and the $z = 1$ $\lambda$ dynamics
$D_{\lambda}(q,i\Omega) = \frac{1}{\kappa_{\theta}
\sqrt{\Omega^{2} + q^{2}}}$, where $\kappa_{\theta}$ is associated
with the compressibility of critical holons.
%
%
%
As discussed in Refs. \cite{Senthil_MIT1,Senthil_MIT2}, the $z =
2$ gauge dynamics does not affect the holon dynamics, preserving
the 3D-XY universality class to allow the universal conductivity
\cite{SF_MI_Original,Cha_Universal_Conductance} at the two
dimensional Mott critical point. {\it This gives a universal
number to $\mathcal{C}_{b}$, resulting in the universal jump for
the Hall coefficient at the Mott critical point.} On the other
hand, if spinons are in the diffusive regime, the gauge dynamics
is given by $D_{a}(\boldsymbol{q},i\Omega) = \frac{1}{\sigma_{f}
|\Omega| + \sigma_{\theta} q}$ with the dc spinon conductivity
$\sigma_{f}$. Such gauge fluctuations will affect the holon
dynamics, and the Mott transition does not belong to the XY
universality class any more. Then, the universal conductivity
depends on the damping coefficient $\sigma_{f}$, thus
$\mathcal{C}_{b}$ becomes a function of disorder and the density
of states of spinons. The Hall coefficient does not become
universal. However, it must be different from the Fermi-liquid
value, giving rise to two-step jumps in the Hall coefficient at
the Mott critical point.
%
%


We claim that the (universal) jump in the Hall conductivity will
not arise from either the DMFT framework or SB-MIT theory. As
revealed in the quantum Boltzman equation approach, the Hall
coefficient does not renormalize due to quantum corrections.
Recalling that the configuration of a Fermi surface is unchanged
at the Mott critical point in these frameworks, the Hall
coefficient of the Mott critical point remains the same as the
Fermi-liquid value, and jumps to an infinite directly in either
the paramagnetic insulator or an antiferromagnetic insulator (Fig.
1).
%
%

One may criticize the assumption for the existence of the
paramagnetic insulating state, referred as a spin liquid phase
\cite{Review_SL}. Although metal-insulator transitions are usually
involved with symmetry breaking,
%
such an insulating phase has been proposed in the DMFT framework
\cite{PM_MI_DMFT} and the gauge theory scenario
\cite{GT_MIT,PM_MI_GT}. In particular, such a state has been
observed by various numerical simulations when geometrically
frustrated lattices are considered to prohibit magnetic ordering
\cite{SL_Simulation}.
%
%
However, the stability of the spin liquid state is not an
important issue in our study. The point is whether electron
fractionalization appears or not at the Mott critical point. If it
occurs, the two-step jumps should be shown in the Hall
coefficient, independent of the nature of an insulating phase.

\section{Conclusion}

In this study we discussed how the spin-charge separation scenario
near two dimensional Mott criticality can be verified from the
Hall coefficient. An essential point is that a novel metallic
state emerges at the Mott critical point, described by critical
dynamics of holon excitations. As a result, such charge
fluctuations contribute to the Hall coefficient additionally on
top of the Fermi liquid value, given by spinon excitations.
Interestingly, the jump of the Hall coefficient turns out to be
universal in the clean limit, given by the 3D-XY universality
class of critical holon dynamics. We claim that the two-step
behavior of the Hall coefficient (Fig. 1) can be regarded as a
fingerprint of electron fractionalization because either the
dynamical mean-field theory or symmetry-breaking metal-insulator
transition allows only one-step jump.

One cautious person may criticize that the behavior of the Hall
coefficient depicted in the right hand side of Fig. 1 and that of
the left hand side do not differ from each other in the
experimental point of view, if the Mott transition occurs really
just at a critical point instead of a critical region. In this
respect the behavior of the Hall coefficient at finite
temperatures may be essential. Since the jump in the Hall
coefficient is given by novel metallicity of holon dynamics, we
are suspecting an interesting scaling behavior in $U/T$ and $H/T$
for the Hall-coefficient jump, where $U$ is a tuning parameter,
here the Hubbard-$U$, and $H$ is externally applied magnetic
field. In particular, the scaling exponent will be given by that
of the 3D-XY universality class in the clean limit. Unfortunately,
we cannot find any previous researches on the scaling behavior of
the Hall coefficient in the superfluid to Mott insulator
transition of the bosonic Hubbard-type model. This will be an
interesting and important future work.

This work was supported by the National Research Foundation of
Korea (NRF) grant funded by the Korea government (MEST) (No.
2011-0074542).

\appendix

\begin{widetext}

\section{To derive the quantum Boltzman equation in the presence of an external magnetic field}

In appendix A we derive the quantum Boltzman equation. This
derivation extends that of Mahan's textbook \cite{Mahan} to the
situation with an external magnetic field. We start from equations
of motion for the lesser Green's function, \bqa &&
[i\partial_{t_1} - H_{0}(r_1,p_1)]G^{<}(x_1,x_2) = \int dx_3
[\Sigma_{t}(x_1,x_3)G^{<}(x_3,x_2) -
\Sigma^{<}(x_1,x_3)G_{\bar{t}}(x_3,x_2)] , \nn &&
[-i\partial_{t_2} - H_{0}(r_2,-p_2)]G^{<}(x_1,x_2) = \int dx_{3}
[G_{t}(x_1,x_3)\Sigma^{<}(x_3,x_2) -
G^{<}(x_1,x_3)\Sigma_{\bar{t}}(x_3,x_2)] .  \eqa The retarded
Green's function remains the same as that in equilibrium
\cite{Mahan}.

First, we focus on the left-hand-side of these equations. These
two equations can be rewritten as follows \bqa &&
[i(\partial_{t_1} - \partial_{t_2}) - H_{1} - H_{2}]G^{<}(x_1,x_2)
= ...  , \nn && [i(\partial_{t_1} +
\partial_{t_2}) - H_{1} + H_{2}]G^{<}(x_1,x_2) = ... ,
\nn && H_{1} = H_{0}(r_1,p_1) , ~~~~~ H_{2} = H_{0}(r_2,-p_2) .
\eqa  The main point is to derive a gauge invariant expression in
the presence of both electric fields and magnetic fields.

Introducing the center of mass $(T,R)$ and relative $(t,r)$
coordinates \bqa && t_{1,2} = T \pm \frac{t}{2} , ~~~~~ r_{1,2} =
R \pm \frac{r}{2} , \eqa
%
we rewrite the covariant derivative of the Hamiltonian as follows
%
%
\bqa && p_{1} + e E_{v} t_1 + \frac{e}{2c} B \times r_{1} = (p +
eE_{v}T + \frac{e}{2c} B \times R) + \frac{1}{2}(P + e E_{v}t +
\frac{e}{2c} B \times r) , \nn && p_{2} - e E_{v} t_2 -
\frac{e}{2c} B \times r_{2} = - (p + eE_{v}T + \frac{e}{2c} B
\times R) + \frac{1}{2}(P + e E_{v}t + \frac{e}{2c} B \times r) ,
\eqa where the applied electric field is decomposed into $E_{v}$
and $E_{s}$ and $p_{1,2} = \pm p + \frac{P}{2}$ is used according
to the coordinate transformation.

Inserting the transformed covariant derivative into the
Hamiltonian part of the equation of motion,
%
we rewrite Eq. (A2) as follows \bqa && 2[i\partial_{t} -
\frac{1}{2m}(p + eE_{v}T + \frac{e}{2c} B \times R)^{2} -
\frac{1}{8m}(P + e E_{v}t + \frac{e}{2c} B \times r)^{2} +
eE_{s}R]G^{<}(x,X) = ... , \nn && [i\partial_{T} - \frac{1}{m}(p +
eE_{v}T + \frac{e}{2c} B \times R)(P + e E_{v}t + \frac{e}{2c} B
\times r) + eE_{s}r] G^{<}(x,X) = ... , \eqa where $\partial_{t_1}
+ \partial_{t_2} = \partial_{T}$ and $\partial_{t_1} -
\partial_{t_2} = 2 \partial_{t}$ are used according to Eq. (A3).

Performing the Fourier transformation, we see
%
%
\bqa && 2[\Omega + e E_{s}R - \frac{1}{2m}(q + eE_{v}T +
\frac{e}{2c} B \times R)^{2} + \frac{1}{8m}(\partial_{R} +
eE_{v}\partial_{\Omega} + \frac{e}{2c} B \times \partial_{q})^{2}]
G^{<}(q,\Omega;R,T) = ... , \nn && i [\partial_{T} + e
E_{s}\partial_{q} + \frac{1}{m}(q + eE_{v}T + \frac{e}{2c} B
\times R)(\partial_{R} + eE_{v}\partial_{\Omega} + \frac{e}{2c} B
\times \partial_{q})] G^{<}(q,\Omega;R,T) = ... . \eqa This
expression is not gauge invariant. In order to make this
expression invariant for the gauge transformation, we perform \bqa
&& \Omega + e E_{s}R \rightarrow \omega , ~~~ q + eE_{v}T +
\frac{e}{2c} B \times R \rightarrow k , \nn &&
\partial_{R} \rightarrow \partial_{R} + e E_{s}
\partial_{\omega} + \frac{e}{2c} B \times \partial_{k} , ~~~
\partial_{T} \rightarrow
\partial_{T} + e E_{v}\partial_{k} . \eqa As a result, we find the
gauge invariant expression in this transformed coordinate, given
by \bqa && [\omega - \epsilon_{k} + \frac{1}{8m}(\partial_{R} + e
E
\partial_{\omega} + \frac{e}{c} B
\times \partial_{k})^{2}] G^{<}(k,\omega;R,T) = ... , \nn && i
[\partial_{T} + v_{k} (\partial_{R} + \frac{e}{c} B \times
\partial_{k}) + e E(\partial_{k} + v_{k}\partial_{\omega})]
G^{<}(k,\omega;R,T) = ...  \eqa

Following Mahan's textbook, the right-hand-side of Eq. (A2) can be
expressed as follows \bqa && [\omega - \epsilon_{k} +
\frac{1}{8m}(\partial_{R} + e E
\partial_{\omega} + \frac{e}{c} B
\times \partial_{k})^{2}] G^{<}(k,\omega;R,T) \nn && = \frac{1}{2}
\int{dz} e^{-iqz} \int{dy} e^{-iqy}
[\Sigma_{t}(y,X+\frac{z}{2})G^{<}(z,X-\frac{y}{2})  -
\Sigma^{<}(y,X+\frac{z}{2})G_{\bar{t}}(z,X-\frac{y}{2}) \nn && +
G_{t}(y,X+\frac{y}{2})\Sigma^{<}(z,X-\frac{y}{2}) -
G^{<}(y,X+\frac{y}{2})\Sigma_{\bar{t}}(z,X-\frac{y}{2}) ]_{GI} ,
\nn && i [\partial_{T} + v_{k} (\partial_{R} + \frac{e}{c} B
\times \partial_{k}) + e E(\partial_{k} + v_{k}\partial_{\omega})]
G^{<}(k,\omega;R,T) \nn && = \int{dz} e^{-iqz} \int{dy} e^{-iqy}
[\Sigma_{t}(y,X+\frac{z}{2})G^{<}(z,X-\frac{y}{2})  -
\Sigma^{<}(y,X+\frac{z}{2})G_{\bar{t}}(z,X-\frac{y}{2}) \nn && -
G_{t}(y,X+\frac{y}{2})\Sigma^{<}(z,X-\frac{y}{2}) +
G^{<}(y,X+\frac{y}{2})\Sigma_{\bar{t}}(z,X-\frac{y}{2}) ]_{GI} ,
\eqa where the subscript "GI" represents gauge invariance, clearer
below.

One can rewrite the self-energy part in a form of the Moyal
product \bqa && I \equiv \int{dz} e^{-iqz} \int{dy} e^{-iqy}
\Sigma (y,X+\frac{z}{2})G (z,X-\frac{y}{2}) =
\Sigma(q,X)e^{\frac{i}{2}[\partial_{q}\partial_{X} -
\partial_{X}\partial_{q}]}G(q,X) , \eqa where the first derivative
in the exponential acts to the front part and the second
derivative does to the behind part.

Near equilibrium, variations with respect to positions are not so
large, allowing us to expand the exponent of derivatives as
follows \bqa && I = \Sigma(q,X)G(q,X) + \frac{i}{2}[
\partial_{q}\Sigma(q,X) \partial_{X}G(q,X) -
\partial_{X}\Sigma(q,X) \partial_{q}G(q,X)] + ...
\nn && \approx \Sigma(q,X)G(q,X) + \frac{i}{2}[\Sigma(q,X),
G(q,X)] . \eqa This is called the gradient expansion. The Poisson
bracket is defined as \bqa && [\Sigma(q,X), G(q,X)] =
\partial_{q}\Sigma(q,X)
\partial_{X}G(q,X) - \partial_{X}\Sigma(q,X) \partial_{q}G(q,X)
\nn && = \partial_{\Omega}\Sigma\partial_{T}G -
\partial_{T}\Sigma\partial_{\Omega}G - (\partial_{q}\Sigma\partial_{R}G -
\partial_{R}\Sigma\partial_{q}G) . \eqa

As discussed in the derivation of the left-hand-side, the above
expression is not gauge invariant in the presence of external
fields. Performing the following transformation for the
right-hand-side \bqa && \Omega + e E_{s}R \rightarrow \omega , ~~~
q + e E_{v}T + \frac{e}{2c} B \times R \rightarrow k , ~~~
\partial_{R} \rightarrow
\partial_{R} + e E_{s} \partial_{\omega} + \frac{e}{2c} B \times \partial_{k} , ~~~ \partial_{T}
\rightarrow \partial_{T} + e E_{v}\partial_{k} , \eqa we write
down the Moyal product in a gauge invariant way \bqa && I =
\Sigma(k,X)e^{\frac{i}{2}[\partial_{k}\partial_{X} -
\partial_{X}\partial_{k} + e E (\partial_{\omega}\partial_{k} -
\partial_{k}\partial_{\omega}) + \frac{e}{c} B \epsilon_{zij} \partial_{k_{i}}\partial_{k_{j}} ]}G(k,X) \nn && \approx
\Sigma(k,X)G(k,X) + \frac{i}{2}[
\partial_{k}\Sigma(k,X) \partial_{X}G(k,X) -
\partial_{X}\Sigma(k,X) \partial_{k}G(k,X)] \nn &&
+ \frac{i}{2}e E [\partial_{\omega}\Sigma(k,X)\partial_{k}G(k,X) -
\partial_{k}\Sigma(k,X)\partial_{\omega}G(k,X)] +
\frac{i}{2} \frac{e}{c} B \epsilon_{zij} \partial_{k_{i}}
\Sigma(k,X) \partial_{k_{j}} G(k,X) . \eqa As a result, we find
two gauge invariant equations of motion for the lesser Green's
function \bqa && [\omega - \epsilon_{k} +
\frac{1}{8m}(\partial_{R} + eE\partial_{\omega} + \frac{e}{c} B
\times \partial_{k})^{2}]G^{<} = \Re\Sigma_{ret} G^{<} +
\Sigma^{<} \Re G_{ret} + \frac{i}{4}[\Sigma^{>},G^{<}] -
\frac{i}{4}[\Sigma^{<},G^{>}] \nn && + \frac{i}{4} e E
[\partial_{\omega}\Sigma^{>}
\partial_{k}G^{<} -
\partial_{k}\Sigma^{<} \partial_{\omega}G^{<}
- \partial_{\omega}\Sigma^{<}\partial_{k}G^{<} +
\partial_{k}\Sigma^{<} \partial_{\omega}G^{>}] \nn && +
\frac{i}{4} \frac{e}{c} B
[\partial_{k_{x}}\Sigma^{>}\partial_{k_{y}}G^{<} -
\partial_{k_{y}}\Sigma^{<}\partial_{k_{x}}G^{<}
- \partial_{k_{x}}\Sigma^{<}\partial_{k_{y}}G^{<} +
\partial_{k_{y}}\Sigma^{<}\partial_{k_{x}}G^{>} ] , \nn &&
i[\partial_{T} + v_{k} (\partial_{R} + \frac{e}{c} B \times
\partial_{k}) + e E \{(1 -
\partial_{\omega}\Re\Sigma_{ret}) \partial_{k} + (v_{k} + \partial_{k} \Re
\Sigma_{ret})\partial_{\omega}\} + \frac{e}{c} B \epsilon_{zij}
\partial_{k_{i}} \Re \Sigma_{ret} \partial_{k_{j}}]
G^{<} \nn && - i e E [\partial_{\omega}\Sigma^{<}\partial_{k}\Re
G_{ret} - \partial_{k}\Sigma^{<} \partial_{\omega} \Re G_{ret} ] -
i \frac{e}{c} B \epsilon_{zij}
\partial_{k_i}\Sigma^{<} \partial_{k_j} \Re G_{ret} =
\Sigma^{>}G^{<} - \Sigma^{<}G^{>} + i [\Re \Sigma_{ret},G^{<}] + i
[\Sigma^{<},\Re G_{ret}] , \nn \eqa where the following relations
are utilized \bqa && G_{t} - G_{\bar{t}} = 2 \Re G_{ret} , ~~~~~
G_{t} + G_{\bar{t}} = G^{<} + G^{>} , \nn && \Sigma_{t} -
\Sigma_{\bar{t}} = 2 \Re \Sigma_{ret} , ~~~~~ \Sigma_{t} +
\Sigma_{\bar{t}} = \Sigma^{<} + \Sigma^{>} . \eqa

Near equilibrium, systems are expected to be in homogeneous
($\partial_{R} G^{<} \rightarrow 0$) and steady ($\partial_{T}
G^{<} \rightarrow 0$) states. Then, the Poisson bracket vanishes
identically. These two equations can be further simplified as
follows \bqa && [\omega - \epsilon_{k} ]G^{<} = \Re\Sigma_{ret}
G^{<} + \Sigma^{<} \Re G_{ret} + \frac{i}{4} e E
[\partial_{\omega}\Sigma^{>}\partial_{k}G^{<} -
\partial_{k}\Sigma^{<} \partial_{\omega}G^{<}
- \partial_{\omega}\Sigma^{<}\partial_{k}G^{<} +
\partial_{k}\Sigma^{<} \partial_{\omega}G^{>}] \nn && +
\frac{i}{4} \frac{e}{c} B
[\partial_{k_{x}}\Sigma^{>}\partial_{k_{y}}G^{<} -
\partial_{k_{y}}\Sigma^{<}\partial_{k_{x}}G^{<}
- \partial_{k_{x}}\Sigma^{<}\partial_{k_{y}}G^{<} +
\partial_{k_{y}}\Sigma^{<}\partial_{k_{x}}G^{>} ] , \nn && \frac{e}{c} (v_{k}+\partial_{k}\Re\Sigma_{ret}) (B \times
\partial_{k}) G^{<} + e E [(1 -
\partial_{\omega}\Sigma_{ret})\partial_{k} +
(v_{k}+\partial_{k}\Re\Sigma_{ret})\partial_{\omega}]G^{<} \nn &&
- e E [\partial_{\omega}\Sigma^{<}\partial_{k}\Re G_{ret} -
\partial_{k}\Sigma^{<} \partial_{\omega} \Re G_{ret} ] -
\frac{e}{c} B \epsilon_{zij}
\partial_{k_i}\Sigma^{<} \partial_{k_j} \Re G_{ret}
= \Sigma^{<} A - 2 \Gamma G^{<} , \eqa where the following
identities of $G^{>} = G^{<} - i A$ and $\Sigma^{>} = \Sigma^{<} -
2 i \Gamma$ are used for the second equation.

The first equation gives rise to the following solution in the
linear response regime \bqa && G^{<}(k,\omega) = i f(\omega)
A(k,\omega) + \mathcal{O}(E^{n};n\geq 1) , \eqa where \bqa &&
A(k,\omega) = \frac{2\Gamma}{(\omega - \epsilon_{k} - \Re
\Sigma_{ret})^{2} + \Gamma^{2}} \eqa is the spectral function with
$\Gamma = - 2 \Im \Sigma_{ret}$.

Inserting this solution into the second equation, we reach the
quantum Boltzman equation \bqa && \frac{e}{c} (v_{k} +
\partial_{k}\Re \Sigma_{ret}) (B \times
\partial_{k}) G^{<} + e E[(v_{k} +
\partial_{k}\Re \Sigma_{ret})\Gamma + (\omega - \epsilon_{k} - \Re
\Sigma_{ret}) \partial_{k} \Gamma] [
\partial_{\omega} f(\omega)] [A(k,\omega)]^{2}
= - i (\Sigma^{<} A - 2 \Gamma G^{<})   . \nn \eqa


\section{To derive $\Lambda_{f(b)}(k,\omega)$ from the quantum
Boltzman equation}

\subsection{To derive $\Lambda_{f}(k,\omega)$}

Inserting the ansatz for the lesser Green's function \bqa &&
G_{f}^{<}(k+q,\omega+\nu) = i A_{f}(k+q,\omega+\nu) \Bigl\{
n_{f}(\omega+\nu) - \Bigl( \frac{\partial
n_{f}(\omega+\nu)}{\partial \omega} \Bigr)
\boldsymbol{v}^{f}_{k+q} \cdot
\boldsymbol{\Lambda}_{f}(k+q,\omega+\nu) \Bigr\} \eqa into the
self-energy expression, we find \bqa && \Sigma_{f}^{<}(k,\omega) =
i \sum_{q} \int_{0}^{\infty} \frac{d\nu}{\pi} \Bigl|
\frac{k\times\hat{q}}{m_{f}} \Bigr|^{2} \Im D_{a}(q,\nu) \nn &&
\Bigl[ \{ n_{b}(\nu) + n_{f}(\omega+\nu) \} A_{f}(k+q,\omega+\nu)
- \{ n_{b}(-\nu) + n_{f}(\omega-\nu) \} A_{f}(k+q,\omega-\nu)
\Bigr] n_{f}(\omega) \nn && + i \sum_{q} \int_{0}^{\infty}
\frac{d\nu}{\pi} \Bigl| \frac{k\times\hat{q}}{m_{f}} \Bigr|^{2}
\Im D_{a}(q,\nu) \Bigl[ \{ n_{b}(\nu) + n_{f}(\omega+\nu) \}
\frac{1 - n_{f}(\omega+\nu)}{1 - n_{f}(\omega)}
A_{f}(k+q,\omega+\nu) \boldsymbol{v}^{f}_{k+q} \cdot
\boldsymbol{\Lambda}_{f}(k+q,\omega+\nu) \nn && - \{ n_{b}(-\nu) +
n_{f}(\omega-\nu) \} \frac{1 - n_{f}(\omega-\nu)}{1 -
n_{f}(\omega)} A_{f}(k+q,\omega-\nu) \boldsymbol{v}^{f}_{k+q}
\cdot \boldsymbol{\Lambda}_{f}(k+q,\omega-\nu) \Bigr] \Bigl(
-\frac{\partial n_{f}(\omega)}{\partial \omega} \Bigr) , \eqa
where the following identities for thermal factors, \bqa &&
\{n_{b}(\nu) + 1\} n_{f}(\omega+\nu) = n_{f}(\omega) \{ n_{b}(\nu)
+ n_{f}(\omega+\nu) \} , ~~~~~ n_{b}(\nu) n_{f}(\omega - \nu) = -
n_{f}(\omega) \{ n_{b}(-\nu) + n_{f}(\omega-\nu) \} \eqa and \bqa
&& \{n_{b}(\nu) + 1\} \Bigl( -\frac{\partial
n_{f}(\omega+\nu)}{\partial \omega} \Bigr) = \{ n_{b}(\nu) +
n_{f}(\omega+\nu) \} \Bigl( -\frac{\partial
n_{f}(\omega)}{\partial \omega} \Bigr) \frac{1 -
n_{f}(\omega+\nu)}{1 - n_{f}(\omega)} , \nn && n_{b}(\nu) \Bigl(
-\frac{\partial n_{f}(\omega-\nu)}{\partial \omega} \Bigr) = - \{
n_{b}(-\nu) + n_{f}(\omega-\nu) \} \Bigl( -\frac{\partial
n_{f}(\omega)}{\partial \omega} \Bigr) \frac{1 -
n_{f}(\omega-\nu)}{1 - n_{f}(\omega)} \eqa are utilized.

Inserting both the lesser Green's function and lesser self-energy
into the quantum Boltzman equation, we obtain the following
expression \bqa && i \frac{e}{c} [
\boldsymbol{v}^{f}_{\boldsymbol{k}} \cdot
(\boldsymbol{\mathcal{B}} \times
\boldsymbol{\partial}_{\boldsymbol{k}})
\boldsymbol{v}^{f}_{\boldsymbol{k}} ] \cdot
\boldsymbol{\Lambda}_{f}(k,\omega) - e \boldsymbol{\mathcal{E}}
\cdot \boldsymbol{v}^{f}_{k} \Gamma_{f}(k,\omega) A_{f}(k,\omega)
= - 2 \Gamma_{f}(k,\omega) \boldsymbol{v}^{f}_{k} \cdot
\boldsymbol{\Lambda}_{f}(k,\omega) \nn && + \sum_{q}
\int_{0}^{\infty} \frac{d\nu}{\pi} \Bigl|
\frac{k\times\hat{q}}{m_{f}} \Bigr|^{2} \Im D_{a}(q,\nu) \Bigl\{
\{ n_{b}(\nu) + n_{f}(\omega+\nu) \} A_{f}(k+q,\omega+\nu)
\boldsymbol{v}^{f}_{k+q} \cdot
\boldsymbol{\Lambda}_{f}(k+q,\omega+\nu) \nn && - \{ n_{b}(-\nu) +
n_{f}(\omega-\nu) \} A_{f}(k+q,\omega-\nu)
\boldsymbol{v}^{f}_{k+q} \cdot
\boldsymbol{\Lambda}_{f}(k+q,\omega-\nu) \Bigr\} , \eqa where \bqa
&& 2 \Gamma_{f}(k,\omega) = \sum_{q} \int_{0}^{\infty}
\frac{d\nu}{\pi} \Bigl| \frac{k\times\hat{q}}{m_{f}} \Bigr|^{2}
\Im D_{a}(q,\nu) \nn && \Bigl\{ \{ n_{b}(\nu) + n_{f}(\omega+\nu)
\} A_{f}(k+q,\omega+\nu) - \{ n_{b}(-\nu) + n_{f}(\omega-\nu) \}
A_{f}(k+q,\omega-\nu) \Bigr\} \eqa is the relaxation rate.

We write down the above expression as follows, introducing $x$ and
$y$ components explicitly \bqa && - i \frac{e}{m_{f} c}
\mathcal{B} \Lambda_{fy}(k,\omega) - [A_{f}(k,\omega)] e
\mathcal{E}_{x} \Gamma_{f}(k,\omega) = - 2 \Gamma_{f}(k,\omega)
\Lambda_{fx}(k,\omega) \nn && + \sum_{q} \int_{0}^{\infty}
\frac{d\nu}{\pi} \Bigl| \frac{k\times\hat{q}}{m_{f}} \Bigr|^{2}
\Im D_{a}(q,\nu) \Bigl\{ \{ n_{b}(\nu) + n_{f}(\omega+\nu) \}
A_{f}(k+q,\omega+\nu) \frac{v_{k+q}^{fx}}{v_{k}^{fx}}
\Lambda_{fx}(k+q,\omega+\nu) \nn && - \{ n_{b}(-\nu) +
n_{f}(\omega-\nu) \} A_{f}(k+q,\omega-\nu)
\frac{v_{k+q}^{fx}}{v_{k}^{fx}} \Lambda_{fx}(k+q,\omega-\nu)
\Bigr\} , \nn && i \frac{e}{m_{f} c} \mathcal{B}
\Lambda_{fx}(k,\omega) - [A_{f}(k,\omega)] e \mathcal{E}_{y}
\Gamma_{f}(k,\omega) = - 2 \Gamma_{f}(k,\omega)
\Lambda_{fy}(k,\omega) \nn && + \sum_{q} \int_{0}^{\infty}
\frac{d\nu}{\pi} \Bigl| \frac{k\times\hat{q}}{m_{f}} \Bigr|^{2}
\Im D_{a}(q,\nu) \Bigl\{ \{ n_{b}(\nu) + n_{f}(\omega+\nu) \}
A_{f}(k+q,\omega+\nu) \frac{v_{k+q}^{fy}}{v_{k}^{fy}}
\Lambda_{fy}(k+q,\omega+\nu) \nn && - \{ n_{b}(-\nu) +
n_{f}(\omega-\nu) \} A_{f}(k+q,\omega-\nu)
\frac{v_{k+q}^{fy}}{v_{k}^{fy}} \Lambda_{fy}(k+q,\omega-\nu)
\Bigr\} . \eqa

Introducing \bqa && \Lambda_{x}(k,\omega) + i
\Lambda_{y}(k,\omega) = \Lambda(k,\omega) , ~~~~~ \mathcal{E}_{x}
+ i \mathcal{E}_{y} = \mathcal{E} , \eqa the quantum Boltzmann
equation becomes \bqa && i \frac{e}{m_{f} c} \mathcal{B}
\Lambda_{f} (k,\omega) - e \mathcal{E} \Gamma_{f}(k,\omega)
A_{f}(k,\omega) = - 2 \Gamma_{f}(k,\omega) \Lambda_{f} (k,\omega)
\nn && + \sum_{q} \int_{0}^{\infty} \frac{d\nu}{\pi} \Bigl|
\frac{k\times\hat{q}}{m_{f}} \Bigr|^{2} \Im D_{a}(q,\nu) \Bigl\{
\{ n_{b}(\nu) + n_{f}(\omega+\nu) \} A_{f}(k+q,\omega+\nu) \cos
\theta \Lambda_{f} (k+q,\omega+\nu) \nn && - \{ n_{b}(-\nu) +
n_{f}(\omega-\nu) \} A_{f}(k+q,\omega-\nu) \cos \theta \Lambda_{f}
(k+q,\omega-\nu) \Bigr\} , \eqa where $\cos \theta$ is introduced
for the velocity ratio.

Considering the Fermi surface, we perform the following
approximation \bqa && \Bigl(1 + i \frac{e \mathcal{B}}{m_{f} c}
\frac{1}{2\Gamma_{f}(k_{F},\omega)} \Bigr) \Lambda_{f}
(k_{F},\omega) = \frac{A_{f}(k_{F},\omega)}{2} e \mathcal{E} +
\frac{1}{2\Gamma_{f}(k_{F},\omega)} \sum_{q} \int_{0}^{\infty}
\frac{d\nu}{\pi} \Bigl| \frac{k_{F}\times\hat{q}}{m_{f}}
\Bigr|^{2} \Im D_{a}(q,\nu) \nn && \Bigl\{ \{ n_{b}(\nu) +
n_{f}(\omega+\nu) \} A_{f}(k_{F}+q,\omega+\nu) - \{ n_{b}(-\nu) +
n_{f}(\omega-\nu) \} A_{f}(k_{F}+q,\omega-\nu) \Bigr\} \cos \theta
\Lambda_{f} (k_{F},\omega) . \eqa As a result, we reach the final
expression for the vertex distribution function \bqa &&
\Lambda_{f}(k_{F},\omega) = \frac{e}{2}
\frac{\tau_{tr}^{f}(k_{F},\omega)}{\tau_{sc}^{f}(k_{F},\omega)}
\frac{A_{f}(k_{F},\omega)}{1 + i \omega_{c}^{f}
\tau_{tr}^{f}(k_{F},\omega)} \mathcal{E} , \eqa where \bqa &&
\frac{1}{\tau_{sc}^{f}(k_{F},\omega)} \equiv 2
\Gamma_{f}(k_{F},\omega) = \frac{N_{F}^{f}}{ 2\pi }
\int{d\xi}\int_{-1}^{1} {d\cos\theta} \int_{0}^{\infty}
\frac{d\nu}{\pi} [v_{F}^{f2}\cos^{2}(\theta/2)] \Im D_{a}(q,\nu)
\nn && \Bigl\{ \{ n_{b}(\nu) + n_{f}(\omega+\nu) \}
A_{f}(k_{F}+q,\omega+\nu) - \{ n_{b}(-\nu) + n_{f}(\omega-\nu) \}
A_{f}(k_{F}+q,\omega-\nu) \Bigr\} \eqa is the relaxation rate and
\bqa && \frac{1}{\tau_{tr}^{f}(k_{F},\omega)} = \frac{N_{F}^{f}}{
2\pi } \int{d\xi}\int_{-1}^{1} {d\cos\theta} \int_{0}^{\infty}
\frac{d\nu}{\pi} [v_{F}^{f2}\cos^{2}(\theta/2)] \Im D_{a}(q,\nu)
\nn && \Bigl\{ \{ n_{b}(\nu) + n_{f}(\omega+\nu) \}
A_{f}(k_{F}+q,\omega+\nu) - \{ n_{b}(-\nu) + n_{f}(\omega-\nu) \}
A_{f}(k_{F}+q,\omega-\nu) \Bigr\} [1 - \cos\theta] \eqa is the
scattering rate associated with transport. Note that there is $1 -
\cos \theta$ factor in the transport time.

The spinon current is given by \bqa && J_{f\mu} = - i
\int\frac{d^{3}k}{(2\pi)^{3}} \frac{k_{\mu}}{m_{f}}
\int\frac{d\omega}{2\pi} G_{f}^{<}(k,\omega) , \eqa resulting in
\bqa && J_{fx} = \int\frac{d^{3}k}{(2\pi)^{3}}
\int\frac{d\omega}{2\pi} \Bigl(- \frac{\partial
n_{f}(\omega)}{\partial \omega} \Bigr) A_{f}(k,\omega) v_{k}^{fx
2} \Lambda_{fx}(k,\omega) , \nn && J_{fy} =
\int\frac{d^{3}k}{(2\pi)^{3}} \int\frac{d\omega}{2\pi} \Bigl(-
\frac{\partial n_{f}(\omega)}{\partial \omega} \Bigr)
A_{f}(k,\omega) v_{k}^{fy 2} \Lambda_{fy}(k,\omega)  .  \eqa Then,
we obtain \bqa && J_{f} = J_{fx} + i J_{fy} = v_{F}^{f2}
\int\frac{d^{3}k}{(2\pi)^{3}} \int\frac{d\omega}{2\pi} \sin^{2}
\theta_{k} \Bigl(- \frac{\partial n_{f}(\omega)}{\partial \omega}
\Bigr) A_{f}(k,\omega) \Lambda_{f} (k,\omega) \nn && = e^{2}
v_{F}^{f2} \int\frac{d^{3}k}{(2\pi)^{3}} \int\frac{d\omega}{2\pi}
\sin^{2} \theta_{k_{F}} \Bigl(- \frac{\partial
n_{f}(\omega)}{\partial \omega} \Bigr) [A_{f}(k_{F},\omega)]^{2}
\frac{\tau_{tr}^{f}(k_{F},\omega)}{\tau_{sc}^{f}(k_{F},\omega)}
\frac{1}{1 + i \omega_{c}^{f} \tau_{tr}^{f}(k_{F},\omega)}
\mathcal{E} \nn && = e^{2} \mathcal{C}_{f} \rho_{F}^{f} v_{F}^{f2}
\frac{\tau_{tr}^{f}(T)}{1 + i \omega_{c}^{f} \tau_{tr}^{f}(T)}
\mathcal{E} , \eqa where \bqa && \mathcal{C}_{f} = \mathcal{C}
\int_{-1}^{1} d \cos \theta_{k_{F}} \sin^{2} \theta_{k_{F}}
\int_{-\infty}^{\infty} d \xi
\frac{1}{\tau_{sc}^{f}(k_{F},\omega)} [A_{f}(\xi,\omega)]^{2} \eqa
with a positive numerical constant $\mathcal{C}$. Note that the
relaxation time in the denominator is cancelled after the $\xi$
integration, making $\mathcal{C}_{f}$ a positive numerical
constant, too. The final expression in Eq. (B16) with Eq. (B17)
coincides with the well known result, justifying our scheme of
approximation.


\subsection{To derive $\Lambda_{b}(k,\omega)$}

One can derive the vertex distribution function for holons in a
similar way of that for spinons. Inserting the lesser Green's
function \bqa && G_{b}^{<}(k+q,\omega+\nu) = i
A_{b}(k+q,\omega+\nu) \Bigl\{ n_{b}(\omega+\nu) - \Bigl(
\frac{\partial n_{b}(\omega+\nu)}{\partial \omega} \Bigr)
\boldsymbol{v}^{b}_{k+q} \cdot
\boldsymbol{\Lambda}_{b}(k+q,\omega+\nu) \Bigr\} \eqa into the
lesser self-energy, we obtain \bqa && \Sigma_{b}^{<}(k,\omega) = i
\sum_{q} \int_{0}^{\infty} \Bigl( \Bigl|
\frac{k\times\hat{q}}{m_{b}} \Bigr|^{2} \Im D_{a}(q,\nu) + \Im
D_{\lambda}(q,\nu) \Bigr) \nn && \Bigl[ \{ n_{b}(\nu) -
n_{b}(\omega+\nu) \} A_{b}(k+q,\omega+\nu) - \{ n_{b}(-\nu) -
n_{b}(\omega-\nu) \} A_{b}(k+q,\omega-\nu) \Bigr] n_{b}(\omega)
\nn && + i \sum_{q} \int_{0}^{\infty} \frac{d\nu}{\pi} \Bigl(
\Bigl| \frac{k\times\hat{q}}{m_{b}} \Bigr|^{2} \Im D_{a}(q,\nu) +
\Im D_{\lambda}(q,\nu) \Bigr) \nn && \Bigl[ \{ n_{b}(\nu) -
n_{b}(\omega+\nu) \} \frac{1 + n_{b}(\omega+\nu)}{1 +
n_{b}(\omega)} A_{b}(k+q,\omega+\nu) \boldsymbol{v}^{b}_{k+q}
\cdot \boldsymbol{\Lambda}_{b}(k+q,\omega+\nu) \nn && - \{
n_{b}(-\nu) - n_{b}(\omega-\nu) \} \frac{1 + n_{b}(\omega-\nu)}{1
+ n_{b}(\omega)} A_{b}(k+q,\omega-\nu) \boldsymbol{v}^{b}_{k+q}
\cdot \boldsymbol{\Lambda}_{b}(k+q,\omega-\nu) \Bigr] \Bigl(
-\frac{\partial n_{b}(\omega)}{\partial \omega} \Bigr) , \eqa
where the following identities, \bqa && \{n_{b}(\nu) + 1\}
n_{b}(\omega+\nu) = n_{b}(\omega) \{ n_{b}(\nu) -
n_{b}(\omega+\nu) \} , ~~~~~ n_{b}(\nu) n_{b}(\omega - \nu) = -
n_{b}(\omega) \{ n_{b}(-\nu)- n_{b}(\omega-\nu) \} \eqa and \bqa
&& \{n_{b}(\nu) + 1\} \Bigl( -\frac{\partial
n_{b}(\omega+\nu)}{\partial \omega} \Bigr) = \{ n_{b}(\nu) -
n_{b}(\omega+\nu) \} \Bigl( -\frac{\partial
n_{b}(\omega)}{\partial \omega} \Bigr) \frac{1 +
n_{b}(\omega+\nu)}{1 + n_{b}(\omega)} , \nn && n_{b}(\nu) \Bigl(
-\frac{\partial n_{b}(\omega-\nu)}{\partial \omega} \Bigr) = - \{
n_{b}(-\nu) - n_{b}(\omega-\nu) \} \Bigl( -\frac{\partial
n_{b}(\omega)}{\partial \omega} \Bigr) \frac{1 +
n_{b}(\omega-\nu)}{1 + n_{b}(\omega)} ,  \eqa are utilized.

Inserting this lesser self-energy and the lesser Green's function
into the quantum Boltzman equation, we obtain \bqa && i
\frac{e}{c} \boldsymbol{v}_{\boldsymbol{k}}^{b} \cdot
[(\boldsymbol{\mathcal{B}} + \boldsymbol{B}) \times
\boldsymbol{\partial}_{\boldsymbol{k}} \{
\boldsymbol{v}_{\boldsymbol{k}}^{b} \cdot
\boldsymbol{\Lambda}_{b}(k,\omega)\}] - e
(\boldsymbol{\mathcal{E}} + \boldsymbol{E}) \cdot
\boldsymbol{v}_{k}^{b} \Gamma_{b}(k,\omega) A_{b}(k,\omega) = - 2
\Gamma_{b}(k,\omega) \boldsymbol{v}_{k}^{b} \cdot
\boldsymbol{\Lambda}_{b}(k,\omega) \nn && + \sum_{q}
\int_{0}^{\infty} \frac{d\nu}{\pi} \Bigl( \Bigl|
\frac{k\times\hat{q}}{m_{b}} \Bigr|^{2} \Im D_{a}(q,\nu) + \Im
D_{\lambda}(q,\nu) \Bigr) \Bigl\{ \{ n_{b}(\nu) -
n_{b}(\omega+\nu) \} A_{b}(k+q,\omega+\nu)
\boldsymbol{v}_{k+q}^{b} \cdot
\boldsymbol{\Lambda}_{b}(k+q,\omega+\nu) \nn && - \{n_{b}(-\nu) -
n_{b}(\omega-\nu) \} A_{b}(k+q,\omega-\nu)
\boldsymbol{v}_{k+q}^{b} \cdot
\boldsymbol{\Lambda}_{b}(k+q,\omega-\nu) \Bigr\} , \eqa where \bqa
&& 2 \Gamma_{b}(k,\omega) = \sum_{q} \int_{0}^{\infty}
\frac{d\nu}{\pi} \Bigl( \Bigl| \frac{k\times\hat{q}}{m_{b}}
\Bigr|^{2} \Im D_{a}(q,\nu) + \Im D_{\lambda}(q,\nu) \Bigr) \nn &&
\Bigl\{ \{ n_{b} (\nu) - n_{b}(\omega+\nu) \}
A_{b}(k+q,\omega+\nu) - \{ n_{b}(-\nu) - n_{b}(\omega-\nu) \}
A_{b}(k+q,\omega-\nu) \Bigr\} \eqa is the relaxation rate.

We write down the above expression as follows, introducing $x$ and
$y$ components explicitly \bqa && - i \frac{e}{m_{b} c}
(\mathcal{B} + B) \Lambda_{b}^{y}(k,\omega) - e (\mathcal{E}_{x} +
E_{x}) \Gamma_{b}(k,\omega) A_{b}(k,\omega) = - 2
\Gamma_{b}(k,\omega) \Lambda_{b}^{x}(k,\omega) \nn && + \sum_{q}
\int_{0}^{\infty} \frac{d\nu}{\pi} \Bigl( \Bigl|
\frac{k\times\hat{q}}{m_{b}} \Bigr|^{2} \Im D_{a}(q,\nu) + \Im
D_{\lambda}(q,\nu) \Bigr) \Bigl\{ \{ n_{b}(\nu) -
n_{b}(\omega+\nu) \} A_{b}(k+q,\omega+\nu) \frac{v_{k+q}^{b
x}}{v_{k}^{b x}} \Lambda_{b}^{x}(k+q,\omega+\nu) \nn && -
\{n_{b}(-\nu) - n_{b}(\omega-\nu) \} A_{b}(k+q,\omega-\nu)
\frac{v_{k+q}^{b x}}{v_{k}^{b x}} \Lambda_{b}^{x}(k+q,\omega-\nu)
\Bigr\} , \nn && i \frac{e}{m_{b} c} (\mathcal{B} + B)
\Lambda_{b}^{x}(k,\omega) - e (\mathcal{E}_{y} + E_{y})
\Gamma_{b}(k,\omega) A_{b}(k,\omega) = - 2 \Gamma_{b}(k,\omega)
\Lambda_{b}^{y}(k,\omega) \nn && + \sum_{q} \int_{0}^{\infty}
\frac{d\nu}{\pi} \Bigl( \Bigl| \frac{k\times\hat{q}}{m_{b}}
\Bigr|^{2} \Im D_{a}(q,\nu) + \Im D_{\lambda}(q,\nu) \Bigr)
\Bigl\{ \{ n_{b}(\nu) - n_{b}(\omega+\nu) \} A_{b}(k+q,\omega+\nu)
\frac{v_{k+q}^{b y}}{v_{k}^{b y}} \Lambda_{b}^{y}(k+q,\omega+\nu)
\nn && - \{n_{b}(-\nu) - n_{b}(\omega-\nu) \}
A_{b}(k+q,\omega-\nu) \frac{v_{k+q}^{b y}}{v_{k}^{b y}}
\Lambda_{b}^{y}(k+q,\omega-\nu) \Bigr\} . \eqa

Introducing \bqa && \Lambda_{b}^{x}(k,\omega) + i
\Lambda_{b}^{y}(k,\omega) = \Lambda_{b}(k,i\omega) , ~~~~~ E_{x} +
i E_{y} = E , \eqa the quantum Boltzman equation becomes \bqa && i
\frac{e}{m_{b} c} (\mathcal{B} + B) \Lambda_{b} (k,\omega) - e
(\mathcal{E} + E ) \Gamma_{b}(k,\omega) A_{b}(k,\omega) = - 2
\Gamma_{b}(k,\omega) \Lambda_{b} (k,\omega) \nn && + \sum_{q}
\int_{0}^{\infty} \frac{d\nu}{\pi} \Bigl( \Bigl|
\frac{k\times\hat{q}}{m_{b}} \Bigr|^{2} \Im D_{a}(q,\nu) + \Im
D_{\lambda}(q,\nu) \Bigr) \Bigl\{ \{ n_{b}(\nu) -
n_{b}(\omega+\nu) \} A_{b}(k+q,\omega+\nu) \cos \theta \Lambda_{b}
(k+q,\omega+\nu) \nn && - \{n_{b}(-\nu) - n_{b}(\omega-\nu) \}
A_{b}(k+q,\omega-\nu) \cos \theta \Lambda_{b} (k+q,\omega-\nu)
\Bigr\} , \eqa where $\cos \theta$ is introduced for the velocity
ratio.

This expression is further simplified, neglecting $q$ and $\nu$ in
the vertex distribution function, \bqa && i \frac{e}{m_{b} c}
(\mathcal{B} + B) \Lambda_{b} (k,\omega) - e (\mathcal{E} + E )
\Gamma_{b}(k,\omega) A_{b}(k,\omega) = - 2 \Gamma_{b}(k,\omega)
\Lambda_{b} (k,\omega) \nn && + \sum_{q} \int_{0}^{\infty}
\frac{d\nu}{\pi} \Bigl( \Bigl| \frac{k\times\hat{q}}{m_{b}}
\Bigr|^{2} \Im D_{a}(q,\nu) + \Im D_{\lambda}(q,\nu) \Bigr)
\Bigl\{ \{ n_{b}(\nu) - n_{b}(\omega+\nu) \} A_{b}(k+q,\omega+\nu)
\nn && - \{n_{b}(-\nu) - n_{b}(\omega-\nu) \}
A_{b}(k+q,\omega-\nu) \Bigr\} \cos \theta \Lambda_{b} (k,\omega) .
\eqa Then, we reach the final expression \bqa &&
\Lambda_{b}(k,\omega) = \frac{e}{2}
\frac{\tau_{tr}^{b}(k,\omega)}{\tau_{sc}^{b}(k,\omega)}
\frac{A_{b}(k,\omega)}{1 + i (\Omega_{b} + \omega_{b})
\tau_{tr}^{b}(k,\omega)} ( \mathcal{E} + E ) , \eqa where \bqa &&
\frac{1}{\tau_{sc}^{b}(k,\omega)} = 2 \Gamma_{b}(k,\omega) =
\frac{1}{ (2\pi)^{2} } \int_{0}^{\infty} d q q^{2} \int_{-1}^{1} d
\cos\theta \int_{0}^{\infty} \frac{d\nu}{\pi} \Bigl(
\frac{k^{2}}{m_{b}^{2}} \cos^{2}(\theta/2) \Im D_{a}(q,\nu) + \Im
D_{\lambda}(q,\nu) \Bigr) \nn && \Bigl\{ \{ n_{b}(\nu) -
n_{b}(\omega+\nu) \} A_{b}(k+q,\omega+\nu) - \{ n_{b}(-\nu) -
n_{b}(\omega-\nu) \} A_{b}(k+q,\omega-\nu) \Bigr\} \eqa is the
relaxation rate and \bqa && \frac{1}{\tau_{tr}^{b}(k,\omega)} =
\frac{1}{ (2\pi)^{2} } \int_{0}^{\infty} d q q^{2} \int_{-1}^{1} d
\cos\theta \int_{0}^{\infty} \frac{d\nu}{\pi} \Bigl(
\frac{k^{2}}{m_{b}^{2}} \cos^{2}(\theta/2) \Im D_{a}(q,\nu) + \Im
D_{\lambda}(q,\nu) \Bigr) \nn && \Bigl\{ \{ n_{b}(\nu) -
n_{b}(\omega+\nu) \} A_{b}(k+q,\omega+\nu) - \{ n_{b}(-\nu) -
n_{b}(\omega-\nu) \} A_{b}(k+q,\omega-\nu) \Bigr\} [1 -
\cos\theta] \eqa is the scattering rate associated with transport.

\section{To derive the longitudinal and Hall conductivities from
the current formulae}

The back-flow constraint results in \bqa &&
\boldsymbol{\mathcal{E}} = - \frac{\mathcal{C}_{b} e^{2}
\frac{\tau_{tr}^{b}(T)}{1 + i \frac{\chi_{f}}{\chi_{f} + \chi_{b}}
\omega_{b} \tau_{tr}^{b}(T)}}{\mathcal{C}_{f} \rho_{F}^{f} e^{2}
v_{F}^{f 2} \frac{\tau_{tr}^{f}(T)}{1 + i \frac{\chi_{b}}{\chi_{f}
+ \chi_{b}} \omega_{f} \tau_{tr}^{f}(T)} + \mathcal{C}_{b} e^{2}
\frac{\tau_{tr}^{b}(T)}{1 + i \frac{\chi_{f}}{\chi_{f} + \chi_{b}}
\omega_{b} \tau_{tr}^{b}(T)}} \boldsymbol{E} . \eqa Then, the
electric current is given by \bqa J_{el} &=& e^{2}
\frac{\mathcal{C}_{f} \mathcal{C}_{b} \rho_{F}^{f} v_{F}^{f 2}
\frac{\tau_{tr}^{f}(T)}{1 + i \frac{\chi_{b}}{\chi_{f} + \chi_{b}}
\omega_{f} \tau_{tr}^{f}(T)} \frac{\tau_{tr}^{b}(T)}{1 + i
\frac{\chi_{f}}{\chi_{f} + \chi_{b}} \omega_{b}
\tau_{tr}^{b}(T)}}{\mathcal{C}_{f} \rho_{F}^{f} v_{F}^{f 2}
\frac{\tau_{tr}^{f}(T)}{1 + i \frac{\chi_{b}}{\chi_{f} + \chi_{b}}
\omega_{f} \tau_{tr}^{f}(T)} + \mathcal{C}_{b}
\frac{\tau_{tr}^{b}(T)}{1 + i \frac{\chi_{f}}{\chi_{f} + \chi_{b}}
\omega_{b} \tau_{tr}^{b}(T)}} \boldsymbol{E} . \eqa

Cooking the above expression as follows \bqa && J_{el} = e^{2}
\frac{\mathcal{C}_{f} \mathcal{C}_{b} \rho_{F}^{f} v_{F}^{f 2}
\tau_{tr}^{f}(T) \tau_{tr}^{b}(T) }{ \Bigl( \mathcal{C}_{f}
\rho_{F}^{f} v_{F}^{f 2} \tau_{tr}^{f}(T) + \mathcal{C}_{b}
\tau_{tr}^{b}(T) \Bigr)^{2} + \Bigl(\mathcal{C}_{f} \rho_{F}^{f}
v_{F}^{f 2} \tau_{tr}^{f}(T) \frac{\chi_{f}}{\chi_{f} + \chi_{b}}
\omega_{b} \tau_{tr}^{b}(T) + \mathcal{C}_{b}
\tau_{tr}^{b}(T)\frac{\chi_{b}}{\chi_{f} + \chi_{b}} \omega_{f}
\tau_{tr}^{f}(T) \Bigr)^{2} } \nn && \Bigl\{ \mathcal{C}_{f}
\rho_{F}^{f} v_{F}^{f 2} \tau_{tr}^{f}(T) + \mathcal{C}_{b}
\tau_{tr}^{b}(T) - i \Bigl(\mathcal{C}_{f} \rho_{F}^{f} v_{F}^{f
2} \tau_{tr}^{f}(T) \frac{\chi_{f}}{\chi_{f} + \chi_{b}}
\omega_{b} \tau_{tr}^{b}(T) + \mathcal{C}_{b}
\tau_{tr}^{b}(T)\frac{\chi_{b}}{\chi_{f} + \chi_{b}} \omega_{f}
\tau_{tr}^{f}(T)\Bigr) \Bigr\} (E_{x} + i E_{y}) , \eqa we obtain
both longitudinal and transverse conductivities, \bqa &&
\sigma_{xx}(T) = e^{2} \frac{\mathcal{C}_{f} \mathcal{C}_{b}
\rho_{F}^{f} v_{F}^{f 2} \tau_{tr}^{f}(T) \tau_{tr}^{b}(T) \Bigl(
\mathcal{C}_{f} \rho_{F}^{f} v_{F}^{f 2} \tau_{tr}^{f}(T) +
\mathcal{C}_{b} \tau_{tr}^{b}(T) \Bigr) }{ \Bigl( \mathcal{C}_{f}
\rho_{F}^{f} v_{F}^{f 2} \tau_{tr}^{f}(T) + \mathcal{C}_{b}
\tau_{tr}^{b}(T) \Bigr)^{2} + \Bigl(\mathcal{C}_{f} \rho_{F}^{f}
v_{F}^{f 2} \tau_{tr}^{f}(T) \frac{\chi_{f}}{\chi_{f} + \chi_{b}}
\omega_{b} \tau_{tr}^{b}(T) + \mathcal{C}_{b}
\tau_{tr}^{b}(T)\frac{\chi_{b}}{\chi_{f} + \chi_{b}} \omega_{f}
\tau_{tr}^{f}(T) \Bigr)^{2} } , \nn && \sigma_{xy}(T) = e^{2}
\frac{\mathcal{C}_{f} \mathcal{C}_{b} \rho_{F}^{f} v_{F}^{f 2}
\tau_{tr}^{f}(T) \tau_{tr}^{b}(T) }{ \Bigl( \mathcal{C}_{f}
\rho_{F}^{f} v_{F}^{f 2} \tau_{tr}^{f}(T) + \mathcal{C}_{b}
\tau_{tr}^{b}(T) \Bigr)^{2} + \Bigl(\mathcal{C}_{f} \rho_{F}^{f}
v_{F}^{f 2} \tau_{tr}^{f}(T) \frac{\chi_{f}}{\chi_{f} + \chi_{b}}
\omega_{b} \tau_{tr}^{b}(T) + \mathcal{C}_{b}
\tau_{tr}^{b}(T)\frac{\chi_{b}}{\chi_{f} + \chi_{b}} \omega_{f}
\tau_{tr}^{f}(T) \Bigr)^{2} } \nn && \Bigl(\mathcal{C}_{f}
\rho_{F}^{f} v_{F}^{f 2} \tau_{tr}^{f}(T) \frac{\chi_{f}}{\chi_{f}
+ \chi_{b}} \omega_{b} \tau_{tr}^{b}(T) + \mathcal{C}_{b}
\tau_{tr}^{b}(T)\frac{\chi_{b}}{\chi_{f} + \chi_{b}} \omega_{f}
\tau_{tr}^{f}(T)\Bigr) , \eqa respectively. These conductivities
give rise to the Hall resistivity of Eq. (12).

\section{To evaluate the holon current}

The holon current is given by \bqa && J_{b} =
\int\frac{d^{3}k}{(2\pi)^{3}} \int\frac{d\omega}{2\pi}
\frac{k^{2}}{m_{b}^{2}} \sin^{2} \theta_{k} \Bigl(- \frac{\partial
n_{b}(\omega)}{\partial \omega} \Bigr) A_{b}(k,\omega) \Lambda_{b}
(k,\omega) . \eqa Inserting the vertex distribution function into
the above, we see \bqa && J_{b} = e^{2}
\int\frac{d^{3}k}{(2\pi)^{3}} \int\frac{d\omega}{2\pi}
\frac{k^{2}}{m_{b}^{2}} \sin^{2} \theta_{k} \Bigl(- \frac{\partial
n_{b}(\omega)}{\partial \omega} \Bigr) [A_{b}(k,\omega)]^{2}
\frac{\tau_{tr}^{b}(k,\omega)}{\tau_{sc}^{b}(k,\omega)} \frac{1}{1
+ i (\Omega_{b} + \omega_{b}) \tau_{tr}^{b}(k,\omega)}
(\boldsymbol{\mathcal{E}} + \boldsymbol{E}) , \eqa where \bqa &&
A_{b}(k,\omega) = \frac{[\tau_{sc}^{b}(k,\omega)]^{-1}}{(\omega -
\xi_{k})^{2} + [\tau_{sc}^{b}(k,\omega)]^{-2}} \eqa is the holon
spectral function.

In the Eliashberg theory the relaxation time is assumed to be
momentum-independent, $\tau_{sc}^{b}(k,\omega) \approx
\tau_{sc}^{b}(\omega)$, where the singular contribution results
from the frequency dependence. Accordingly, the transport time is
assumed to depend on frequency only.

Resorting to this assumption, we obtain the current formula \bqa
&& J_{b} = \mathcal{C}_{\theta} \frac{e^{2}}{m_{b}^{2}}
\int\frac{d\omega}{2\pi} \int_{0}^{\infty} d k k^{4} \Bigl(-
\frac{\partial n_{b}(\omega)}{\partial \omega} \Bigr) \Bigl(
\frac{[\tau_{sc}^{b}(\omega)]^{-1}}{(\omega - \xi_{k})^{2} +
[\tau_{sc}^{b}(\omega)]^{-2}} \Bigr)^{2}
\frac{\tau_{tr}^{b}(\omega)}{\tau_{sc}^{b}(\omega)} \frac{1}{1 + i
(\Omega_{b} + \omega_{b}) \tau_{tr}^{b}(\omega)}
(\boldsymbol{\mathcal{E}} + \boldsymbol{E})
%
%
\nn && = \mathcal{C}_{\theta} \frac{e^{2}}{m_{b}^{2}}
\int\frac{d\omega}{2\pi} \Bigl(- \frac{\partial
n_{b}(\omega)}{\partial \omega} \Bigr) \int_{(\omega + \mu)
\tau_{sc}^{b}(\omega)}^{\infty} d y \Bigl(y
[\tau_{sc}^{b}(\omega)]^{-1} + \omega + \mu \Bigr)^{3/2} \Bigl(
\frac{1}{y^{2} + 1} \Bigr)^{2} \frac{\tau_{tr}^{b}(\omega)}{1 + i
(\Omega_{b} + \omega_{b}) \tau_{tr}^{b}(\omega)}
(\boldsymbol{\mathcal{E}} + \boldsymbol{E}) \nn && =
\mathcal{C}_{\theta} \frac{e^{2}}{m_{b}^{2}}
\int\frac{d\omega}{2\pi} \Bigl(- \frac{\partial
n_{b}(\omega)}{\partial \omega} \Bigr) \int_{1}^{\infty} d z (z +
1 )^{3/2} \frac{(\omega + \mu)^{5/2} \tau_{sc}^{b}(\omega)
}{[(\omega+\mu)^{2} \{ \tau_{sc}^{b}(\omega) \}^{2} z^{2}+1]^{2}}
\frac{\tau_{tr}^{b}(\omega)}{1 + i (\Omega_{b} + \omega_{b})
\tau_{tr}^{b}(\omega)} (\boldsymbol{\mathcal{E}} + \boldsymbol{E})
\nn && \approx \mathcal{C}_{\theta} \frac{e^{2}}{m_{b}^{2}}
\int_{1}^{\infty} d z (z + 1 )^{3/2} \frac{(T + \mu)^{5/2}
\tau_{sc}^{b}(T) }{[(T+\mu)^{2} \{ \tau_{sc}^{b}(T) \}^{2}
z^{2}+1]^{2}} \frac{\tau_{tr}^{b}(T)}{1 + i (\Omega_{b} +
\omega_{b}) \tau_{tr}^{b}(T)} (\boldsymbol{\mathcal{E}} +
\boldsymbol{E}) \nn && \approx e^{2} \mathcal{C}_{b}
\frac{\tau_{tr}^{b}(T)}{1 + i (\Omega_{b} + \omega_{b})
\tau_{tr}^{b}(T)} (\boldsymbol{\mathcal{E}} + \boldsymbol{E}) ,
\eqa where $\mathcal{C}_{\theta}$ is a positive numerical constant
associated with the angular integration, and $\mathcal{C}_{b}$ is
given by \bqa && \mathcal{C}_{b} \approx
\frac{\mathcal{C}_{\theta}}{m_{b}^{2}} \Bigl[ \Bigl\{
\int_{1}^{(T+\mu)^{-1} \{ \tau_{sc}^{b}(T) \}^{-1}} d z (z + 1
)^{3/2} (T + \mu)^{5/2} \tau_{sc}^{b}(T) \nn && +
\int_{(T+\mu)^{-1} \{ \tau_{sc}^{b}(T) \}^{-1}}^{\infty} d z
\frac{(z + 1 )^{3/2}}{z^{4}} \frac{1}{ (T+\mu)^{3/2} \{
\tau_{sc}^{b}(T) \}^{3}} \Bigr\} \Theta\Bigl((T+\mu)^{-1} \{
\tau_{sc}^{b}(T) \}^{-1} - 1\Bigr) \nn && + \int_{1}^{\infty} d z
\frac{(z + 1 )^{3/2}}{z^{4}} \frac{1}{ (T+\mu)^{3/2} \{
\tau_{sc}^{b}(T) \}^{3}} \Theta\Bigl(1 - (T+\mu)^{-1} \{
\tau_{sc}^{b}(T) \}^{-1}\Bigr) \Bigr] \nn && \approx
\frac{\mathcal{C}_{\theta}}{m_{b}^{2}} \Bigl[ \Bigl\{
\frac{16}{15} [\tau_{sc}^{b}(T)]^{-3/2} - \frac{4}{5}
(T+\mu)^{5/2} \tau_{sc}^{b}(T) \Bigr\} \Theta\Bigl((T+\mu)^{-1} \{
\tau_{sc}^{b}(T) \}^{-1} - 1\Bigr) \nn && + \frac{2}{3} \frac{1}{
(T+\mu)^{3/2} \{ \tau_{sc}^{b}(T) \}^{3}} \Theta\Bigl(1 -
(T+\mu)^{-1} \{ \tau_{sc}^{b}(T) \}^{-1}\Bigr) \Bigr]  . \eqa This
expression tells us that $\mathcal{C}_{b}$ vanishes as $T$ goes to
zero in the paramagnetic Mott insulator because the presence of an
excitation gap ($\mu \not= 0$) implies irrelevance of self-energy
corrections in the $T \rightarrow 0$ limit. The only contribution
resulting from the second term dies out in this limit. On the
other hand, the first term contributes to $\mathcal{C}_{b}$ at the
Mott critical point ($\mu = 0$). The self-energy correction due to
$\lambda$ fluctuations is given by $\sim \int d^{3} q q^{-1}
|\boldsymbol{k}+\boldsymbol{q}|^{-2} \sim \int_{-1}^{1} d \cos
\theta \int_{0}^{\infty} d q q^{2} q^{-1} (k^{2} + q^{2} + 2 k q
\cos \theta)^{-1} = \int_{-1}^{1} d \cos \theta \int_{0}^{\infty}
d x x (1 + x^{2} + 2 x \cos \theta)^{-1}$, causing
$[\tau_{sc}^{b}(T)]^{-1} = const.$ in one time and two space
dimensions. As a result, we obtain a finite value for
$\mathcal{C}_{b}$. However, it turns out to vanish in three
spatial dimensions, allowing only one-step jump for the Hall
coefficient.

\end{widetext}


\begin{thebibliography}{9}
\bibitem{Lambda_Dynamics} In order to describe the Mott transition
from a spin liquid state to a Fermi liquid phase, it is convenient
to resort to the description of the nonlinear $\sigma$ model.
Performing $e^{i\theta_{\boldsymbol{x} }} \rightarrow
\psi_{\boldsymbol{x} }$ with the unimodular constraint
$|\psi_{\boldsymbol{x} }|^{2} = 1$, one can rewrite the effective
Lagrangian Eq. (2) as follows \bqa && Z = \int D
\psi_{\boldsymbol{x} } D f_{\boldsymbol{x} \sigma} D
\lambda_{\boldsymbol{x}} D \varphi_{\boldsymbol{x} } D
\boldsymbol{a}_{\boldsymbol{x}} e^{ - \int_{0}^{\beta} d \tau \int
d^{2} \boldsymbol{r} {\cal L}} , \nn && {\cal L} =
f_{\boldsymbol{x} \sigma}^{\dagger} (\partial_{\tau} - \mu - i
\varphi_{\boldsymbol{x} }) f_{\boldsymbol{x} \sigma} +
\frac{1}{2m_{f}} |(\boldsymbol{\nabla} - i
\boldsymbol{a}_{\boldsymbol{x}}) f_{\boldsymbol{x} \sigma}|^{2}
\nn && - \frac{1}{2 u} (\psi^{*}_{\boldsymbol{x} }
\partial_{\tau} \psi_{\boldsymbol{x}
} - i \varphi_{\boldsymbol{x} })^{2} + \frac{1}{2m_{b}}
|(\boldsymbol{\nabla} - i
\boldsymbol{a}_{\boldsymbol{x}})\psi_{\boldsymbol{x} }|^{2} \nn &&
+ i \lambda_{\boldsymbol{x}} (|\psi_{\boldsymbol{x} }|^{2} - 1) +
\frac{1}{4g^{2}} f_{\mu\nu} f_{\mu\nu} . \nonumber \eqa
$\lambda_{\boldsymbol{x}}$ is introduced to keep the rotor
constraint, where $\lambda_{\boldsymbol{x}}$ fluctuations play an
important role in the self-energy correction of the
$\psi_{\boldsymbol{x}}$ field.
\end{thebibliography}
\end{document}